\documentclass[12pt]{article}
\usepackage{amsmath,ZaZ}
\usepackage{graphicx,color}

\def\TC{T_{\!\sss C\,}}
\def\TR{T_{\!\sss R\,}}

 \SfTitles %section titles more modest
 \allowdisplaybreaks
\begin{document}
 \vglue2mm
 %\vfill
 \begin{center}
{\Large\sf\bfseries\boldmath
  A Classical Switched $LC/LR$ Circuit\\[-1mm]
  Modeling the Quantum Zeno and Anti-Zeno Effects\\[-1mm]
 }{~}\\*[2mm]
  %\vfill
{\sf\bfseries T.\,H\"{u}bsch$^*$,
               and
              V.\,Pankovi\'c$^\dag$
}\\*[0mm]
{\small\it
  $^*$%
      Department of Physics \&\ Astronomy,\\[-1mm]
      Howard University, Washington, DC 20059, USA
%  \\[-4pt] {\tt  thubsch@howard.edu}
  \\[0mm]
  $^\dag$%
      Department of Physics, Faculty of Natural Sciences\\[-1mm]
      University of Novi Sad, 21\,000~Novi Sad, Serbia
%  \\[-4pt] {\tt  thubsch@howard.edu}
 }\\[5mm]
 \vglue2mm
 %\vfill
{\sf\bfseries ABSTRACT}\\[3mm]
\parbox{152mm}{\small\addtolength{\baselineskip}{-2pt}\parindent=2pc\noindent
Generalizing a recent analysis, we model the quantum Zeno and anti-Zeno effects with a quickly switched, ideal $LC/LR$ circuit, in the limiting case of it alternating very many times between its short $LC$ and even shorter $LR$ regime. If the initial current is arranged to be negligible, the capacitor turns out never to discharge. If the initial current is made dominant, this capacitor discharges, and faster than the exponential decay to which it limits.  The existence and simplicity of these phases in such a rudimentary model indicates that the corresponding effects are ubiquitous throughout physics, both quantum and classical.
 In turn, the parameter space of this model contains intermediate phases in which the circuit exhibits behaviors that seem to foreshadow other quantum effects of some interest.
}
\end{center}
 \vglue2mm
 %\vfill
\noindent PACS: 03.65.Ta, 42.50.Lc
%\clearpage
\setcounter{page}{1}
\section{Introduction}
 \label{s:1}
Early analyses of the quantum Zeno effect\cite{rM+S} aimed to initiate a reevaluation of the collapse paradigm of quantum measurement. However, it was soon shown that neither the measurement paradigm nor the type of the initial state (pure or mixed) affects this effect; see for example Ref.\cite{rDz90}. Furthermore, the quantum Zeno effect is also consistent with measurement as a spontaneous (non-dynamical) superposition breaking\cite{rPPK,rPHPK}.
 Subsequent work foreshadowed\cite{rNNP-QZ} and then formulated more precisely\cite{rK+G,rLRQaZ,rFP-QZ,rMPS,rMP-DHO} several similar effects dubbed interchangeably the quantum anti-Zeno, inverse-Zeno, or Heraclites effect; see Ref.\cite{rFP-QZMat} for a recent comprehensive review.
 Recent analysis\cite{rVP-LC} implies that, far from being quantum or even quasi-classical in nature, these effects can actually also occur in purely classical physics.

Herein, we extend this analysis, establishing a correspondence of sorts between the quantum effects and their classical analogues. We also demonstrate that the specification (controllable parameters) of the classical systems permit a transition from the ``Zeno phase'' into the ``anti-Zeno phase,'' and explore their features. By way of this correspondence, certain features of these classical models may well have interesting quantum-mechanical interpretations, relevant to conceptual understanding of these effects, but also to certain applications.

As one of the most familiar toy-models, we consider an initially charged ideal capacitor ($C$) first discharging through an ideal inductor ($L$) for a short period of time. The inductor is then swiftly disconnected from the capacitor and shunted through an ideal resistor ($R$). Finally, the circuit is toggled between the two regimes very quickly.
 Straightforward analysis then shows that, for certain choices of the circuit parameters, the so switched $LC/LR$ circuit never discharges. That is, this circuit exhibits a Zeno effect, and provides a faithful, manifest, and easily constructed classical model of the quantum Zeno effect\cite{rM+S}.
 In turn, for other choices of the circuit parameters, this switched $LC/LR$ circuit does discharge|and faster than the exponential discharge to which it limits. This then offers an equally handy classical model of the quantum anti-Zeno effect\cite{rK+G}.

The simplicity of this switched $LC/LR$ circuit model and its analysis implies the intrinsic simplicity of the effects being modeled. Furthermore, it is not difficult to envision many other analogous classical systems, with the switching mechanism corresponding to frequent interruptions. All such systems then turn out to exhibit the corresponding analogues of the Zeno and anti-Zeno effect, indicating that these two effects, far from being peculiarities of the quantum nature of Nature and the measurement paradigm, are ubiquitous throughout all of physics, quantum and classical.

\section{The Quantum Zeno and Quantum anti-Zeno Effects}
 \label{s:2}
The quantum Zeno effect was first formulated theoretically by Misra and Sudarshan as the quantum Zeno paradox\cite{rM+S}, but is by now an experimentally verified phenomenon\cite{rIHBW,rFGMR} and has become standard textbook material\cite{rEM,rLEB,rDG-QM,rESA}.

Within the standard formalism of quantum mechanics\cite{rJvN} (see also more recent texts, such as\cite{rEM,rLEB,rDG-QM,rESA}), a quantum system may change its state in two essentially different ways. One is the unitary, superposition-conserving and deterministic quantum mechanical dynamical evolution, generated by the Hamiltonian and specified by the Schr\"odinger equation. This applies during time intervals of arbitrary length but during which no measurement process occurs and the system may be regarded as isolated. Following von Neumann's postulate, the other way in which a state can change is the so-called collapse, in which superposition is broken probabilistically by a measurement process, and which is inherently an interaction with an external agent\ft{It has been argued that inclusion of all interacting elements without {\em\/reductio ad absurdum\/} implies a nonlinearity in the dynamics of quantum mechanics that breaks the principle of superposition\cite{rHTQM}.}. This collapse (measurement) requires a finite time interval determined by the Heisenberg indeterminacy relations. Nevertheless, formally and without loss of generality, the collapse may be considered as being instantaneous.

Misra and Sudarshan\cite{rM+S} considered an unstable quantum system with the total Hamiltonian $H$, in its initial quantum state $\ket{i}$. This initial state evolves, during some short time interval $[0,t]$, into the final state $\ket{f} = e^{-iHt/\hbar}\ket{i}$, which is a superposition of the undecayed initial quantum state $\ket{i}$, and the decayed state $\ket{d}$. The final superposition may be corresponded, roughly speaking, to an oscillation between the decayed and the undecayed quantum states. For a short time interval, however, this final state can be approximated by its Taylor expansion to second order:
\begin{equation}
 \ket{f} \approx \Big(\,1 - \frac{iHt}{\hbar}- \frac12\frac{H^2t^2}{\hbar^2} \,\Big)\ket{i}.
 \label{e:1}
\end{equation}
Attempting to measure lack of decay at the time $t>0$ on the quantum system in the final state will thus detect the undecayed state $\ket{i}$ with the quantum-mechanical probability:
\begin{align}
 P_{i\to i}(t)&\Defl\Vev{i|e^{-iHt/\hbar}|i}
                \approx \big(\,1 - (t/\t_i)^2 \,\big),
 \label{e:2}
\intertext{where}
 \quad\t_i
  &\Defl\inv{\hbar}(\triangle_iH)=\inv{\hbar}\sqrt{\vev{i|H^2|i}-\vev{i|H|i}^2}
\end{align}
is the characteristic time of the system in the state $\ket{i}$.
Eq.\eq{e:2} manifestly contains no terms linear in $t$. For this reason, \Eq{e:1} represents the minimal non-trivial approximation of $\ket{f}$.

Now subdivide the small time interval $[0,t]$ into $n$ sub-intervals of the same duration, $\frac{t}{n}$ where $n$ is a natural number, and assume that a decay measurement is realized at the end of each of these time sub-intervals. Since all measurement is (formally) instantaneous, the time spent in measurement is for all practical purposes of measure zero in the interval $[0,t]$.
 Then, the probability that a quantum system will remain undecayed after the whole time interval, by the the time $t$ equals:
\begin{equation}
 P^{\sss(n)}_{i\to i}(t)
  = \Big[\,1 - \Big(\frac{t}{n\t_i}\Big)^2\,\Big]^n
  \approx 1 - \frac1n\Big(\frac{t}{\t_i}\Big)^2
 \label{e:3}
\end{equation}
In the limit $n\to\infty$, this probability clearly tends toward 1, implying that the frequent measurements prevent the unstable quantum system from decaying. This is the well-publicized result: ``a watched pot never boils,'' metaphorically speaking.

The foregoing seems to indicate a paradox: although non-trivial quantum-mechanical dynamical evolution takes place during the whole time interval $[0,t]$, moment by moment, there is nevertheless effectively no dynamical evolution by the end of whole time interval. In fact however, there is no actual paradox: On one hand, all theoretical predictions are obtained strictly adhering to the standard quantum mechanical formalism. On the other, the results are in excellent agreement with experiments\cite{rIHBW,rFGMR}. So, a paradox exists only in the eyes of a na{\"\ii}ve-intuitive beholder, supposing that a measurement (\ie, the interaction of the system with a measuring apparatus) has the form of the simple dynamical evolution.

On the flip-side, Kaulakys and Gontis formulated theoretically the quantum anti-Zeno effect \cite{rK+G}, foreshadowed in the general result of Ref.\cite{rNNP-QZ}, which is mathematically not as simple; see the recent review\cite{rFP-QZMat}. Fortunately, we will not need its details, and suffice it here simply to cite Kaulakys and Gontis: ``{\sl We show that repetitive measurements
 on a multilevel system with quantum suppression of classical
 chaos results in delocalization of the superposition of state,         
 and restoration of chaotic dynamics. Since this effect is the
 reverse of the quantum Zeno effect we call this phenomenon
 the {\it\/quantum anti-Zeno effect\/}.}'' Similar effects were discussed in Ref.\cite{rLRQaZ,rFP-QZ,rMPS,rMP-DHO}, and we refer the Reader to these sources for the details.

\section{Zeno and Anti-Zeno Effects in a Classical $LC$ Circuit}
 \label{s:3}
Herein, we discuss a classical model of the quantum effects described in the previous section.

First, consider a well-known ideal $LC$ circuit with no thermal resistance: an ideal inductor of inductivity $L$ connected in series to a charged, ideal capacitor of capacitance $C$, and an ideal switch closing the circuit. The equation governing this circuit|when closed|is:
\begin{equation}
 \frac{\rd^2\,q(t)}{\rd t^2} = -\w^2\, q(t),\qquad
 \w=\frac1{\sqrt{LC}},
 \label{e:4}
\end{equation}
Its general solution is:
\begin{equation}
 q(t) = q_0\cos(\w t) + \frac{i_0}{\w}\sin(\w t),
 \label{e:LC}
\end{equation}
where $q_0=q(0)$ is the initial electrical charge and $i_0=i(0)=\dot{q}(0)$ is the initial electric current, controlled independently. Suffice it to assume that the capacitor had been charged initially, so $q_0>0$, and that it can but discharge upon completing the circuit through the inductor, so $i_0\leq0$.

The system then has two characteristic time-scales: the ``intrinsic'' one, $\t_\w\Defl\w^{-1}$, and one that is determined purely by initial/boundary conditions: $\t_i\Defl\big|\frac{q_0}{i_0}\big|$. For any sufficiently short time ($t\ll\t_\w,\t_i$) from closing the circuit, the assumption $q_0>0$ implies that $q(t)>0$ and the general solution is well approximated by 
\begin{equation}
 q(t) = q_0 \Big(\,1 + \frac{i_0}{q_0}t - \inv2\w^2t^2\Big).
 \label{e:6}
\end{equation}
The formal analogy of this (indeed very basic!) result with the formula\eq{e:1} is pivotal in our analysis. Furthermore, together with $i_0\leq0$, the positivity of $q(t)$ for such short periods of time then implies:
\begin{alignat}{3}
 -\Big|\frac{i_0}{q_0}\Big|\,t~&>~\inv2\w^2t^2-1,\qquad&\To\qquad
 \frac{t}{\t_i}~&<~1-\inv2\Big(\frac{t}{\t_\w}\Big)^2.
\intertext{Probing the limiting values, we substitute $t\to\t_\w$ and find:}
 \frac{\t_\w}{\t_i}~&<~\inv2,\qquad&
 |i_0|~&<~\inv2|q_0|\,\w.
 \label{e:i<q/2w}
\end{alignat}
Since $\t_i>2\t_\w$, it would have been inconsistent to substitute $t\to\t_i$ as that would have violated the $t\ll\t_\w,\t_i$ assumption; indeed,
\begin{equation}
  \frac{t}{\t_i}<1-\inv2\Big(\frac{t}{\t_\w}\Big)^2\quad\tooo{t\to\t_i}\quad
  1<1-\inv2\Big(\frac{\t_i}{\t_\w}\Big)^2\quad\To\quad
  0<-\inv2\Big(\frac{\t_i}{\t_\w}\Big)^2
\end{equation}
is impossible to satisfy.

\begin{figure}[htbp]
 \begin{center}
  \begin{picture}(80,40)
   \put(-7,-10){\includegraphics[height=50mm]{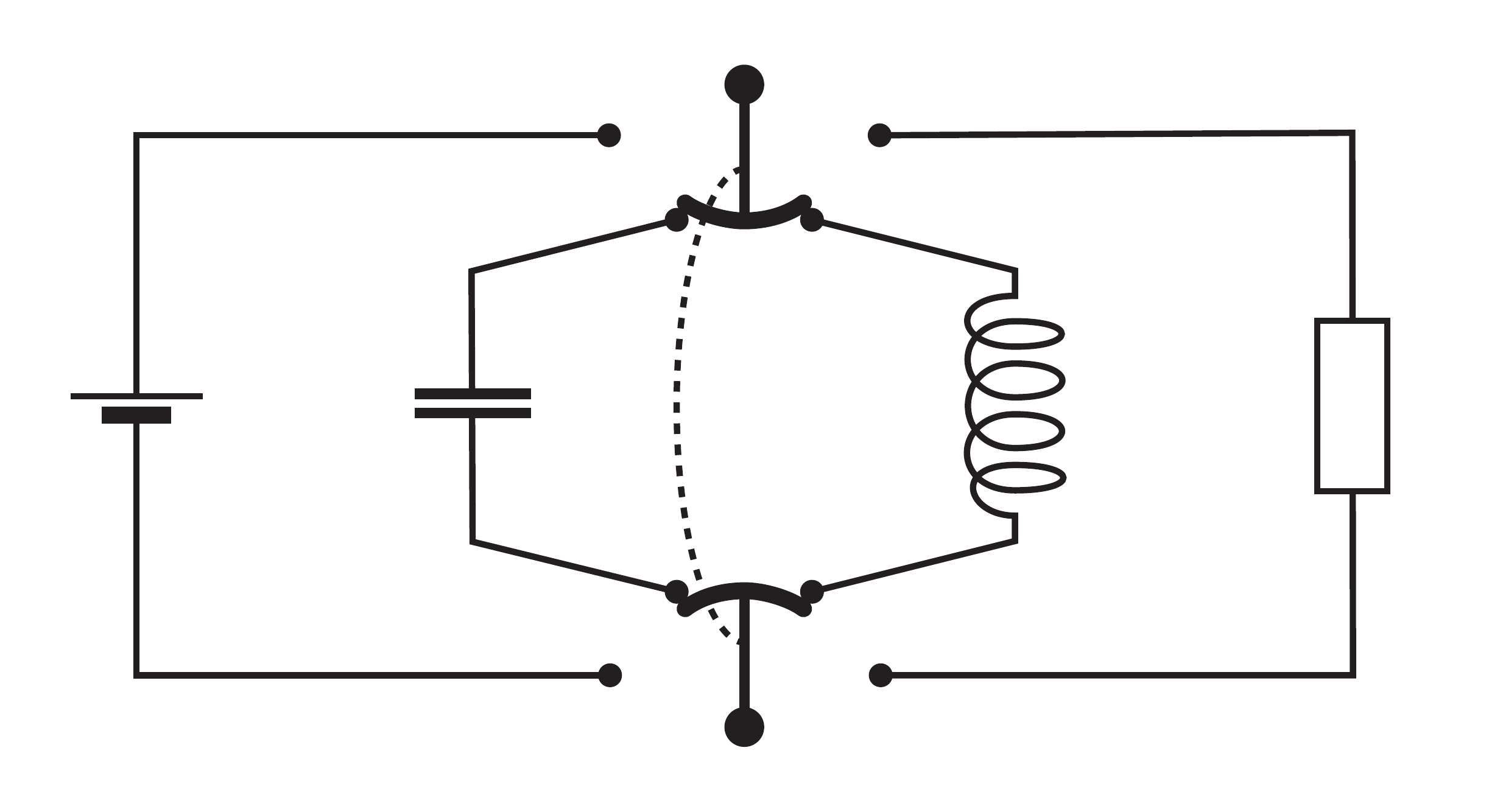}}
   \put(3,9){\large$\cal E$}
   \put(24,9){\large$C$}
   \put(48,9){\large$L$}
   \put(68,9){\large$R$}
   \put(31,27){{}\llap{\footnotesize\sc prepare}}
   \put(37,22){\hbox{\hss\footnotesize\sc on\hss}}
   \put(47,27){\rlap{\footnotesize\sc off}{}}
  \end{picture}
 \end{center}
 \caption{The ideal switched LC/RL circuit. The dotted line indicates that the two physical switches are perfectly synchronized and we refer to this tandem as ``the switch.''}
 \label{f:1}
\end{figure}
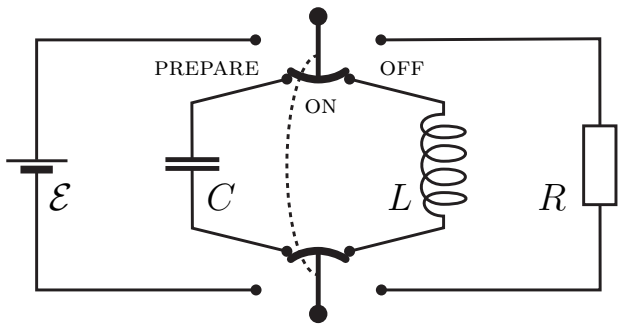

Next, we equip this $LC$ circuit with an ideal switching mechanism, sketched in Fig.~\ref{f:1},
which may also be used to prepare the circuit: in the `{\sc prepare}' position, the capacitor is charged up to the desired initial value, $q_0$. Independently, an initial and stationary current $i_0$ may be established through the inductor by a separate current generator, controlled by the same switch but not shown in the diagram to avoid clutter. All circuit elements are considered ideal, so the double switch will be assumed to make and break all connections practically instantaneously.

When the initial values are prepared, the switch is toggled `{\sc on}' at $t=0$, and the $LC$ circuit behaves as described above. But, at the time $t=\TC$, this switch very quickly toggles `{\sc off},' disconnecting the inductor from the capacitor and simultaneously shunting the inductor through an ideal resistor of resistance $R$; see Fig.~\ref{f:1}.
The resulting closed $LR$ circuit is governed by the equation:
\begin{equation}
 ÐL \frac{\rd\, i_{\sss R}(t)}{\rd t} = R\,i_{\sss R}(t),\qquad
 i_{\sss R}(t)=i_{\sss R}(\TC)\,e^{-\frac{R}{L}(t-T_C)}.
\end{equation}
Since the current through the inductor is routed through the resistor at the time of switching `{\sc off},' we have that $i(\TC)=i_{\sss R}(\TC)$, and so
\begin{equation}
 i_{\sss R} (\TC)=-\big(q_0\,\w^2\,\TC + |i_0|\big)
  ~=~-\Big(\frac{q_0\,\TC}{LC}+|i_0|\Big) < 0.
\end{equation}

We may now choose a brief subsequent time $t=(\TC{+}\TR)$ with $\TR\ll \TC$, such that 
\begin{equation}
 \Big(i_{\sss R}(\TC+\TR)=i_{\sss R}(\TC)\,e^{-\frac{R}{L}(T_R)}\Big)
 ~=~\big(i_0 = - |i_0|\big),
 \label{e:12}
\end{equation}
by selecting $R$:
\begin{equation}
 R = \frac{L}{\TR}\,\ln\Big(\frac{q_0}{|i_0|}\,\w^2\,\TC + 1\Big)
   = \frac{L}{\TR}\,\ln\Big(\frac{q_0\,\TC}{|i_0|\,LC} + 1\Big).
 \label{e:13}
\end{equation}
This ensures that, in the brief period of time $\TR$ after the switch was toggled {\sc off}, the electrical current $i_{\sss R}(t)$ is reset to the $t=0$ initial value\ft{The limiting case $i_0\to0$ implies that $R\to\infty$, so that the inductor is simply disconnected from the capacitor during the {\sc off} regime.}; that is, $i_{\sss R}(\TC{+}\TR)=i_0$.

On the other hand, during $t\in[\TC,(\TC{+}\TR)]$, there is no change in the electrical charge on the capacitor since it was disconnected from the circuit after the switch was toggled {\sc off}.

The above two regimes having completed the switching cycle at $t=(\TC{+}\TR)$, the switch toggles instantaneously from {\sc off} back {\sc on}, and the inductor is again connected in series to the capacitor, and not the resistor. The system is again the $LC$ circuit described by \Eqs{e:4}{e:6}|modeling \Eqs{e:1}{e:3}|with the equivalent initial current condition, but now starting from a somewhat changed initial value of the charge: The final solution of this dynamical equation in the small subsequent time interval $[(\TC{+}\TR),(2\TC+\TR)]$ is:
\begin{equation}
 q(t) = q_0 \big(1-\inv2\w^2\TC^2-\frac{|i_0|}{q_0}\TC\big)
             \Big(1-\inv2\w^2t^2-
                  \frac{|i_0|}{q_0(1-\frac{|i_0|}{q_0}\TC)}t-\frac12\w^2\TC^2\Big) > 0,
\end{equation}
where positivity implies that
\begin{equation}
 q(t) \leq q_0\Big(1-\inv2\w^2\TC^2-\frac{|i_0|}{q_0}\TC\Big)^2.
\end{equation}
After this, the switch goes {\sc off} for another time period of $\TR$, the current resets to $i_0$, the switch goes back {\sc on}, and the cycle repeats.

Consider finally a short time interval $[0,T]$, with $T<\t_\w$, subdivide it into a large number of sub-intervals $[0,\frac{T}{N}]$, and switch the circuit {\sc on-off} once within each sub-interval. That is, arrange so that $(\TC{+}\TR)=\frac{T}{N}$, and the circuit goes through $N$ {\sc on-off} cycles in the short time interval $[0,T]$. For sufficiently many $(N\gg1)$ subdivisions, the condition $\frac{T}{N}=(\TC{+}\TR)\ll\t_\w=\w^{-1}$ is easy to achieve and the conditions of the above analysis are satisfied. Furthermore, since $\TR\ll\TC$, it follows that $\frac{T}{N}=(\TC{+}\TR)\approx \TC$.
 By a simple induction of the preceding results,
\begin{equation}
 q(T) \leq
 q_0 \Big[1-\inv2\w^2\Big(\frac{T}{N}\Big)^2-\frac{|i_0|}{q_0}\Big(\frac{T}{N}\Big)\Big]^N.
 \label{e:16}
\end{equation}

Now, when
\begin{equation}
 \inv2\w^2\Big(\frac{T}{N}\Big)^2 \gg \frac{|i_0|}{q_0}\frac{T}{N},
  \qquad\text{\ie,}\qquad
  |i_0| \ll \inv2 q_0\,\w^2\frac{T}{N}=\inv2\frac{q_0}{LC}\frac{T}{N},
 \label{e:i0-}
\end{equation}
\Eq{e:16} may be approximated as
\begin{equation}
 q(t) \leq q_0\Big[1-\inv2\frac{\w^2T^2}{N^2}\Big]^N  \tooo{~N\to\infty~} q_0.
 \label{e:CLZ}
\end{equation}
This implies that the capacitor remains undischarged and that the electric current does not flow at all during the time interval $[0,T]$, even though the electric current can flow at any isolated moment within the large interval. Metaphorically, the finger meddling inside a toaster never gets burned.

Evidently, the result\eq{e:CLZ} is the hallmark of a classical Zeno effect, perfectly analogous to the quantum Zeno effect|together with the preparation stage, which is essential for the quantum effect\cite{rNNP-QZ}. It is in full agreement with classical electrodynamics (electronics) and represents a paradox only to the na{\"\ii}ve intuition. The situation may be understood on examining Fig.~\ref{f:2}:
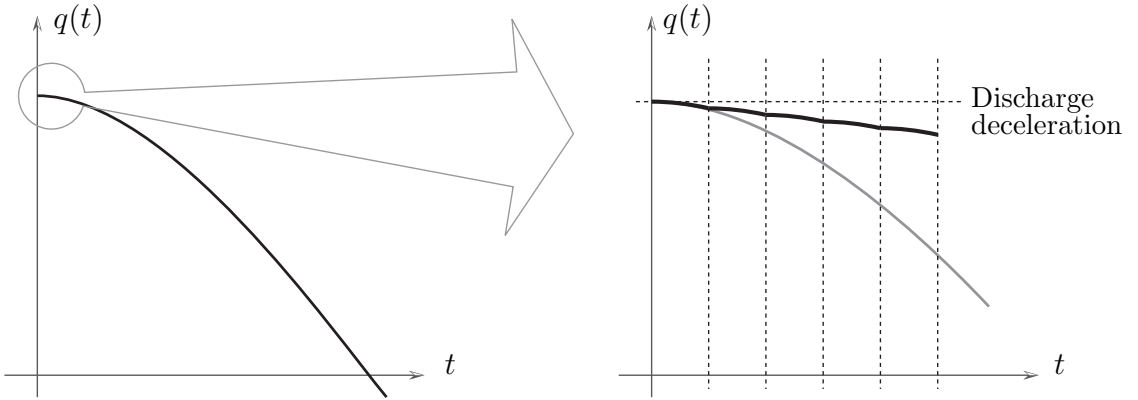
\begin{figure}[htbp]
 \begin{center}
  \begin{picture}(140,50)
   \put(0,-10){\includegraphics[width=140mm]{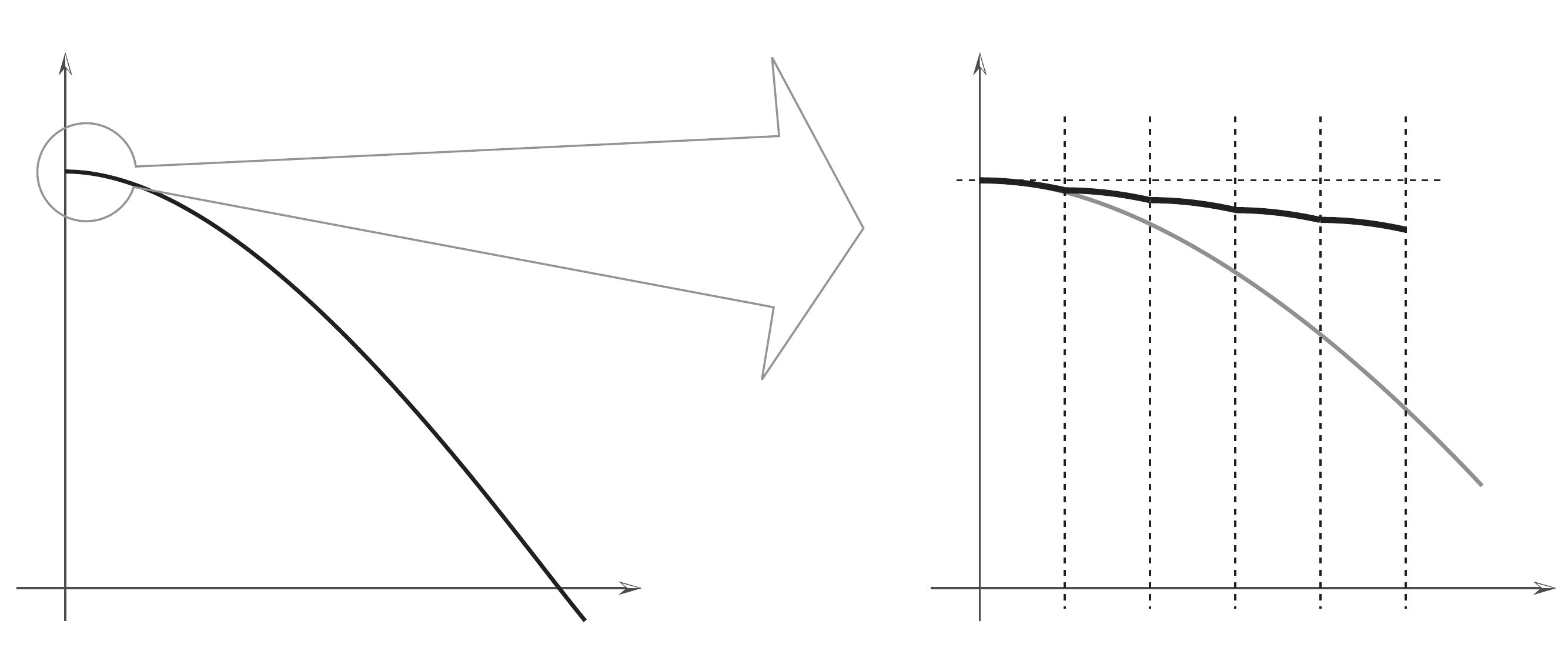}}
   \put(60,-4){$t$}
   \put(8,42){$q(t)$}
   \put(141,-4){$t$}
   \put(89,42){$q(t)$}
   \put(130,30){\parbox{25mm}{\raggedright\small\baselineskip=10pt
                              Discharge deceleration}}
  \end{picture}
 \end{center}
 \caption{A sketch of the effect that the frequent switching has on the switched $LC/RL$ circuit (dark, concatenated curve on the right), as compared with the un-switched $LC$ circuit (dark curve on the left, light gray on the right).}
 \label{f:2}
\end{figure}
the $\cos$-like discharging of the capacitor in an un-switched $LC$ circuit is seen to be replaced here by an iterative concatenation of copies of the initial, very short and nearly flat segment. This leads to a discharging of the capacitor at a pace that is much slower than what happens in the un-switched $LC$ circuit, in which the capacitor is discharged fully by $t=\inv4\t_\w$. In the $N\to\infty$ limit of indefinitely frequent switching, the capacitor does not discharge at all.
\ping

In the opposite case, \ie, when\ft{The perhaps contrary-looking conditions\eq{e:i0+} and\eq{e:i<q/2w} turn out the be perfectly consistent: combining them into $\inv2 q_0\,\w^2\frac{T}{N}\ll|i_0|<\inv2q_0\w$ merely states that $\w\frac{T}{N}\ll1$, \ie, that $\frac{T}{N}\ll\t_\w$ as was assumed. Indeed, we have the hierarchy of time-scales: $\frac{T}{N}\ll\t_\w<2\t_\w<\t_i$.}
\begin{equation}
 \inv2\w^2\Big(\frac{T}{N}\Big)^2 \ll \frac{|i_0|}{q_0}\frac{T}{N},
  \qquad\text{\ie,}\qquad
  |i_0| \gg \inv2 q_0\,\w^2\frac{T}{N}=\inv2\frac{q_0\,T}{N\,LC},
 \label{e:i0+}
\end{equation}
\Eq{e:16} may be approximated as
\begin{equation}
 q(t) \leq q_0\Big[1-\frac{|i_0|}{q_0}\frac{T}{N}\Big]^N,
  \qquad\text{and}\qquad
 \lim_{N\to\infty} q(t) = q_0\,e^{-\frac{|i_0|}{q_0}t},\quad\text{for}~t\in[0,T].
 \label{e:exp}
\end{equation}
tending approximately toward an exponential decay (rather than oscillation!) of the electric charge in the capacitor. Notice however that the electric discharging ends up being faster than the exponential curve to which it limits. Fig.~\ref{f:3}:
\begin{figure}[htbp]
 \begin{center}
  \begin{picture}(140,50)
   \put(0,-10){\includegraphics[width=140mm]{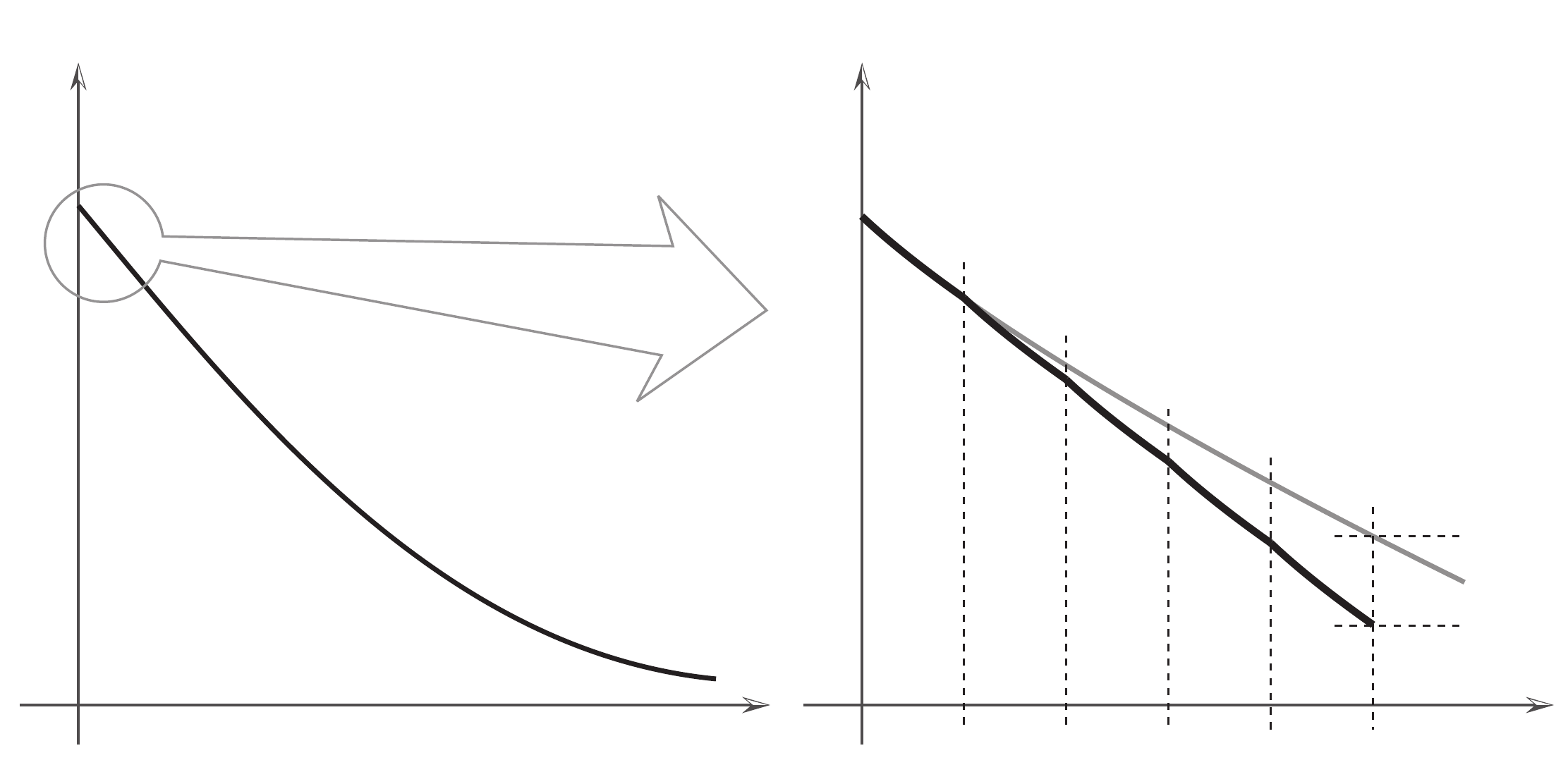}}
   \put(133,7){\parbox{25mm}{\raggedright\small\baselineskip=9pt
                              Discharge acceleration}}
  \end{picture}
 \end{center}
 \caption{A sketch of the effect that the frequent switching and a non-negligible initial current have on the switched $LC/RL$ circuit; this graph is virtually identical to the one reported in Ref.\protect\cite{rLRQaZ}.}
 \label{f:3}
\end{figure}
shows the limiting exponential curve (black on the left, grey on the right) and the finite-$\frac{T}{N}$ curve approximating the tangent with the slope $i_0$ (black on the right).
This result evidently models the similar effects discussed in Refs.\cite{rK+G,rLRQaZ,rFP-QZ}, all dubbed the ``anti-Zeno,'' ``inverse-Zeno'' or ``Heraclites'' effect, emphasizing the relative acceleration as compared to the exponential decay. Metaphorically, the finger meddling inside a toaster gets burned before the toaster heats up.
 In turn, the described dynamics is in full agreement with classical electrodynamics (electronics) and represents a paradox only to the na{\"\ii}ve intuition.
\ping

There are well-known analogies between the classical $LC$ circuit, the classical mechanical linear harmonic oscillators of various kinds: swings, springs, \etc, and so-called small oscillations in the most general Hamilton-Jacobi theory in classical mechanics and field theory|to all of which we refer as ``{\sc lho}.'' This implies that any {\sc lho} with an appropriate and momentarily acting ``meddling mechanism'' that corresponds to the ideal switch with the shunting resistor described above will similarly exhibit a classical Zeno or anti-Zeno effect. In turn, this perfectly corresponds to the general results of Ref.\cite{rNNP-QZ} and also to the thermodynamically influenced quantum {\sc lho} as a system which also exhibits the quantum Zeno and anti-Zeno effects as discussed in Refs.\cite{rMPS,rMP-DHO}; see below. 

Put simply, we consider an ideal (undampened) classical harmonic oscillator that evolves classically during a small time interval $[0,\TC]$ with some initial ``position'' and initial ``velocity.''. In the next, much smaller time interval $[\TC,(\TC{+}\TR)]$ with $\TR\ll\TC$, a specific non-conservative interaction between the oscillator and a meddling mechanism occurs. As soon as this momentary meddling interaction ends, at the time $t=(\TC{+}\TR)=\frac{T}{N}$, the oscillator resumes its temporarily stopped evolution, but with the ``velocity'' restored to its initial, $t=0$ value. This cycle is then repeated many ($N\gg1$) times  during a fixed, and still short time interval $T<\t_\w$, where $\t_\w$ is the characteristic time-scale of the oscillator, \ie, the shortest such scale in complex oscillators.

Such a frequently switched oscillator then exhibits the classical Zeno effect if the initial velocity (current) is negligible|as specified by the inequality\eq{e:i0-}, and the classical anti-Zeno effect if the initial velocity (current) is dominant|as specified by the inequality\eq{e:i0+}.

Finally, note that the resetting of the electric current (during the shunting through the resistor), \ie, the resetting in the ``velocity'' of the oscillator (owing to the non-conservative though momentary meddling interactions) represent a transitions from the dynamical (typically kinetic) energy into thermal energy, in perfect agreement with Joule's law.

\section{Zeno and Anti-Zeno Phases and Conclusions}
 \label{s:4}
Reviewing the previous discussion, we can determine a sequence of criteria, which imply the emergence of the Zeno or the anti-Zeno effect in any domain of physics (classical mechanics, classical electrodynamics, classical thermodynamics, quantum mechanics, \etc):

\begin{description}
  \item[\boldmath$Z_0$:] The Zeno and the anti-Zeno effect occur in physical systems which have two sub-systems; let $\t$ denote the smallest of the characteristic time-scales of the system.

  \item[\boldmath$Z_1$:] During a short period of time $t\in[0,\TC]$ with $\TC\ll\t$, an observable, $\cal R$, of the first subsystem changes through dynamical evolution (the first process) as ${\cal R}_0\to{\cal R}_0(1-at-bt^2)$. Here $a$ and $b$ are time-independent characteristic parameters of the system.

  \item[\boldmath$Z_2$:] During an even shorter subsequent period of time $t\in[\TC,\TC{+}\TR]$, a non-conservative momentary|formally instantaneous|meddling interaction (the second process) between the first and second sub-system changes the observable ${\cal R}_0(1-at-bt^2)\to{\cal R}_0'(1-a't-bt^2)$, where ${\cal R}'_0$ and $a'>a$ are the ``renormalized'' values of these time-independent characteristic parameters. 

  \item[\boldmath$Z_4$:] The system switches between the two processes a large ($N\gg1$) though finite number of times, such that $\TC{+}\TR=\frac{T}{N}$ and $T<\t$.

  \item[\boldmath$Z_5$:]  For $a'\ll b\frac{T}{N}$ the system exhibits the Zeno effect; for $a'\gg b\frac{T}{N}$, it exhibits the anti-Zeno effect.
\end{description}

In most concrete examples, $a,b$ are continuously changeable|and controllable|variables. Also, the length of the time period $\frac{T}{N}$, \ie, the number of subdivisions $N$, and the relative length of $\TC$ {\em\/vs.\/} $\TR$ are free parameters of the system and the switched process. In the concrete case of the switched $LC/LR$ circuit, the choice of the circuit elements $C,L$|with $R$ determined by\eq{e:13}|the initial conditions $q_0,i_0$ and the choice of $\TC,\TR,N$ parametrize the behavior of the switched circuit. Although the 7-parameter space $(C,L,q_0,i_0;\TC,\TR,N)$ may well be redundant\ft{For example, the four values $C,L,q_0,i_0$ determine the two characteristic time constants $\t_\w,\t_i$ and the Zeno and anti-Zeno effects only dependent on these two. This indicates a possible redundancy in the parametrization, \ie, that the true parameter space is less than 7-dimensional.}, it is clear that there exists a nontrivial continuous parameter space for the switched $LC/LR$ circuit. As described above, there are two interesting regions in this parameter space, in one of which the circuit exhibits the Zeno effect, this then being the ``Zeno phase,'' whereas in another region the circuit exhibits the anti-Zeno effect, this then being the ``anti-Zeno phase.''

It should be clear that this does not (by far) exhaust the phase diagram: Comparing the conditions\eq{e:i0-} and\eq{e:i0+}, we realize that the initial current $i_0$ may be assigned values that continuously interpolate between\eq{e:i0-} and\eq{e:i0+}. The correspondingly interpolating behaviors of the switched $LC/LR$ circuit are then expected to interpolate between the Zeno and the anti-Zeno effect|and will likely be affected by different choices of $N$: if nothing else, as the relative ratios of $\frac{T}{N},\t_\w$ and $\t_i$ take on rational values, one expects various resonance effects to emerge.

Whereas the mathematically ideal $N\to\infty$ limiting cases do come in three universality classes owing to the fact that
\begin{equation}
 \lim_{N\to\infty}\Big[\,1+\Big(\frac{x}{N}\Big)^\d\,\Big]^N =
  \left\{\begin{array}{l@{\quad\text{for}~}l}
           1 & \d>1,\\ e^x & \d=1,\\ \infty & \d<1,
         \end{array}\right.
 \label{e:3Cases}
\end{equation}
the finite-$N$|and so more realistic|cases are likely to exhibit a richer structure and set of options. It would then seem to follow that the region of the parameter space interpolating between the Zeno and the anti-Zeno effect will not have a simple demarcation line between the two phases, but possibly an intermediate region with several novel effects. The precise details of this intermediate region are expected to depend on the details of the specific system and the (possibly uneven) subdivision of the time interval $[0,T]$, and so must be explored on a case-by-case basis.

In turn, each of the behaviors described in Section~\ref{s:3} and sketched in Figs.~\ref{f:1} and~\ref{f:2} has a perfect quantum-mechanical analogue and there is a formal correspondence between each quantum result presented in section~\ref{s:2} with its analogue, derived for the classical switched $LC/LR$ circuit in section~\ref{s:3}. Based on this formal correspondence, we conclude that the other, interpolating phases of the classical switched $LC/LR$ circuit should have their corresponding|and perhaps novel|quantum analogues, and we hope to return to a more detailed mapping of this phase diagram at a later opportunity, comparing with the numerous cases discussed in Ref.\cite{rNNP-QZ} and their possible generalizations.

Suffice it here to draw the Reader's attention to the interesting {\em\/qualitative change\/} in the switched $LC/LR$ circuit|controlled by the value of $i_0$: As $i_0$ is initially chosen to be negligible, the circuit exhibits the Zeno effect: the slowing of the discharge as shown in Fig.~\ref{f:1}. Note that each finite segment in Fig.~\ref{f:1}, for $N$ large but finite, does retain the very small but non-vanishing and negative $O(t^2)$ curvature. Indeed, the well-known exact result for the un-switched $LC$ circuit\eq{e:LC} is easily seen to have a {\em\/negative\/} curvature near $t=0$ and arbitrary values of $i_0$ as long as $q_0>0$.
 By contrast, when a non-negligible $i_0<0$ is arranged, Fig.~\ref{f:2} shows that none of the {\em\/finite\/} segments nor their concatenation has any curvature, but limits to the exponential decay in the $N\to\infty$ limit|which in turn has a {\em\/positive\/} $O(t^2)$ curvature.

It then follows that the frequent switching (meddling) radically changes the dynamical evolution of the circuit, and the character of the change may be controlled by arranging the value of $i_0$: Depending on $i_0$, the switched $LC/LR$ circuit either ``remembers'' its characteristic curvature and oscillatory nature, or ``forgets'' it and in the limit acquires an opposite value for exponential evanescence. In view of the correspondence between this switched $LC/LR$ circuit and quantum systems exhibiting the quantum Zeno or anti-Zeno effects, it is intriguing to speculate about possibility of analogously controllable effects of ``quantum memory/senility'' and the interpolation between them.

Finally, we note that the above conclusions are very general: In all of classical physics, for all observables $\cal R$ (functions over the phase space) for which a non-conservative meddling interaction of the kind described in section~\ref{s:3} may be introduced, we have the following general behavior:
\begin{equation}
 \begin{tabular}{ll}
  if $\d{\cal R}(t)\sim O(t^\a)$ with $\a>1$, &there exists an associated Zeno effect;\\
  if $\d{\cal R}(t)\sim O(t)$, &there exists an associated anti-Zeno effect.
 \end{tabular}
 \label{e:R}
\end{equation}
Since the analogous quantum effects are by now well established (theoretically and experimentally), the foregoing implies that these effects exist independently of the ``quantum-ness'' or ``classical-ness'' of the given process. In fact, Ehrenfest's and the Feynman-Hellman theorem and their various generalizations imply that the quantum theory will have to reproduce these classical effects, and so will have to in fact also have their quantum versions\ft{Usually, the quantum Zeno and anti-Zeno effects are discussed in terms of (non-)transition probabilities. The analogous must, however, hold more generally for matrix elements since transition probabilities {\em\/are\/} (squares of) certain particular matrix elements, such as\eq{e:2}, and the classical limit of these can exhibit the Zeno or the anti-Zeno effect as discussed in section~\ref{s:4}. Alternatively, the quantum version of\eq{e:R} may be derived by considering the evolution of operators in the Heisenberg or the interaction picture, and then connecting to the above classical results {\em\/via\/} Ehrenfest- and Feynman-Hellman-like theorems.}.
 These effects|and presumably then also any other effects interpolating between the two as described above|are universal in all of physics. Owing to the result\eq{e:3Cases}, the ideal $N\to\infty$ limit would reduce the phase diagram to only the Zeno and the anti-Zeno phase, with a clear demarcation line. The physically realistic $1\ll N<\infty$ cases are then clearly more interesting in that they may harbor the emergence of intermediate phases.

In conclusion, let us summarize as follows: In this note, we have shown that classical systems such as a switched $LC/LR$ circuit may be arranged to exhibit the Zeno and the anti-Zeno effects. By the well-known analogy, this also corresponds to the classical linear harmonic oscillator equipped with a corresponding meddling mechanism|and its ``small oscillation'' generalizations within all of classical physics. This then permits us to provide a general sequence of criteria for occurrence of the Zeno or anti-Zeno effect within any field of physics. In turn, a quantum physics translation of the above-hinted phase diagram indicates that the emergence of the Zeno and the anti-Zeno effects are also in full agreement with the interpretation of the quantum measurement as a spontaneous superposition breaking phase transition\cite{rPPK,rPHPK}.

\bigskip\paragraph{\bfseries Acknowledgments:}
TH is a visiting professor at the Physics Department of the Faculty of Natural Sciences of the University of Novi Sad, Serbia, and wishes to thank for the recurring hospitality and resources. TH\ is indebted to the generous support of the Department of
Energy through the grant DE-FG02-94ER-40854.
\vfill\clearpage
\small
\bibliographystyle{elsart-numX}
\bibliography{Refs}
\end{document}